\documentclass[pdflatex,sn-mathphys-num]{sn-jnl}

\usepackage{graphicx}%
\usepackage{multirow}%
\usepackage{amsmath,amssymb,amsfonts}%
\usepackage{amsthm}%
\usepackage{mathrsfs}%
\usepackage[title]{appendix}%
\usepackage{xcolor}%
\usepackage{textcomp}%
\usepackage{manyfoot}%
\usepackage{booktabs}%
\usepackage{algorithm}%
\usepackage{algorithmicx}%
\usepackage{algpseudocode}%
\usepackage{listings}%

\usepackage[final]{changes}

\begin{document}

\title{FACT: \textbf{F}oundation Model for \textbf{A}ssessing \textbf{C}ancer \textbf{T}issue Margins with Mass Spectrometry}


\author*[1]{\fnm{Mohammad} \sur{Farahmand}}\email{m.farahmand@queensu.ca}
\author[1]{\fnm{Amoon} \sur{Jamzad}}
\author[2]{\fnm{Fahimeh} \sur{Fooladgar}}
\author[1]{\fnm{Laura} \sur{Connolly}}
\author[1]{\fnm{Martin} \sur{Kaufmann}}
\author[1]{\fnm{Kevin Yi Mi} \sur{Ren}}
\author[1]{\fnm{John} \sur{Rudan}}
\author[1]{\fnm{Doug} \sur{McKay}}
\author[1]{\fnm{Gabor} \sur{Fichtinger}}
\author[1]{\fnm{Parvin} \sur{Mousavi}}

\affil*[1]{\orgname{Queen's University}, \orgaddress{\city{Kingston}, \state{ON}, \country{Canada}}}
\affil*[2]{\orgname{The University of British Columbia}, \orgaddress{\city{Vancouver}, \state{BC}, \country{Canada}}}


\abstract{
    \textbf{Purpose:} Accurately classifying tissue margins during cancer surgeries is crucial for ensuring complete tumor removal. Rapid Evaporative Ionization Mass Spectrometry (REIMS), a tool for real-time intraoperative margin assessment, generates spectra that requires machine learning models to support clinical decision-making. However, the scarcity of labeled data in surgical contexts presents a significant challenge. This study is the first to develop a foundation model tailored specifically for REIMS data, addressing this limitation and advancing real-time surgical margin assessment.
    \textbf{Methods:} We propose \textbf{FACT}, a Foundation model for Assessing Cancer Tissue margins. FACT is an adaptation of a foundation model originally designed for text-audio association, pretrained using our proposed supervised contrastive approach based on triplet loss. An ablation study is performed to compare our proposed model against other models and pretraining methods.
    \textbf{Results:} Our proposed model significantly improves the classification performance, achieving state-of-the-art performance with an AUROC of \(82.4\% \pm 0.8\). The results demonstrate the advantage of our proposed pretraining method and selected backbone over the self-supervised and semi-supervised baselines and alternative models.
    \textbf{Conclusion:} Our findings demonstrate that foundation models, adapted and pretrained using our novel approach, can effectively classify REIMS data even with limited labeled examples. This highlights the viability of foundation models for enhancing real-time surgical margin assessment, particularly in data-scarce clinical environments.
}


\maketitle


\section{Introduction}

Basal cell carcinoma (BCC) is among the most common skin cancers, especially prevalent in the neck and face \cite{manoli2020real}. While BCC is less likely to metastasize, its invasive nature can make complete tumor resection challenging, particularly in cases where the tumor deeply infiltrates healthy surrounding tissue. Surgical resection remains the primary treatment, where achieving negative margins—indicating total tumor removal—is crucial for minimizing recurrence. 

Rapid Evaporative Ionization Mass Spectrometry (REIMS) has emerged as a powerful tool for intraoperative margin assessment \cite{balog2013intraoperative}. REIMS can provide real-time metabolic profiles of tissue by analyzing the surgical smoke generated and collected by iKnife \cite{santilli2020perioperative}. Due to the sheer complexity and the size of REIMS data, effectively utilizing this data for tissue classification is only possible through advanced machine learning techniques. While prior studies have explored the use of deep learning for accurately distinguishing between cancerous and non-cancerous tissue \cite{fooladgar2022uncertainty,connolly2024imspect}, significant challenges remain. Chief among these is the scarcity of labeled data, as annotating mass spectrometry data requires input from expert pathologists. This process is laborious, expensive, and destructive, limiting the scalability of such approaches. In addition, similarities in the metabolic profiles of BCC and certain skin layers frequently lead to high false positive rates. These challenges pose a significant barrier to broader clinical application of REIMS data.

Foundation models are an emerging paradigm in deep learning that offer a promising solution to the challenge of labeled data scarcity in fields like mass spectrometry. These models are large-scale networks pretrained on vast, diverse datasets, allowing them to capture generalizable representations that are transferable across different tasks. By leveraging the general knowledge encoded in these models, they can be fine-tuned for specific downstream applications with significantly fewer labeled examples \cite{awais2023foundational,liang2024foundation,zhou2023comprehensive}. However, in practice, the effectiveness of a foundation model depends on its pretraining data being similar to the target domain. For example, while models like Segment Anything Model (SAM) \cite{kirillov2023segment}—a foundation model for image segmentation that has been pretrained on a large dataset of diverse natural images—show promising results on medical images, domain-specific adaptations such as MedSAM \cite{ma2024segment} demonstrates superior performance on medical images thanks to its specialized pretraining. Currently, there are no  foundation models tailored to REIMS data, which presents a gap in the field that must be addressed to fully leverage the benefits of foundation models in this context.

To address this gap, we present \textbf{FACT}, a Foundation model for Assessing Cancer Tissue margins, the first of its kind to the best of our knowledge. FACT is an adaptation of Contrastive Language-Audio Pretraining (CLAP) \cite{wu2023large}, a state-of-the-art foundation model originally designed for associating text and audio. The spectral characteristics of REIMS data, such as short-time variations in frequency patterns and sudden fluctuations in intensities, share notable similarities with audio Mel-spectrograms, the audio modality that CLAP works with, making CLAP a suitable backbone for our model. Foundation models are generally pretrained using self-supervised or semi-supervised methods, to make use of abundant unlabeled data without having to annotate them. However, in this work, we propose a supervised pretraining method, a contrastive learning approach based on triplet loss. We rationalize that after transfer learning from another modality, foundation models can be adapted with a small but high quality labeled dataset. Our proposed method ensures that samples from the same class (cancerous or benign) are closer in the projected embedding space, while samples from differing classes are as far apart as possible. Our method emphasizes \emph{hard} negatives, spectra that are too similar in the embedding space to those of the opposing class. This results in more robust embeddings, less prone to lead to false positives, improving the results of our primary objective, accurately classifying cancer margins.

We compare CLAP with two other foundation models \cite{radford2021learning,bushuiev2024emergence} we adapt for REIMS data. We also compare our proposed pretraining approach with vastly adopted self-supervised \cite{chen2020simple} and semi-supervised \cite{sohn2020fixmatch} alternatives. Despite only having access to a labeled dataset consisting of 682 spectra, our supervised pretraining approach achieves superior results in classification accuracy than the alternatives, which make use of our substantially larger sets of unlabeled data. Our method demonstrates statistically significant improvements over previous approaches, achieving an AUROC of \(82.4\% \pm 0.8\) and a balanced accuracy of \(77.5\% \pm 1.9\). These results affirm that supervised learning remains a viable strategy in this context, especially when leveraging foundation models with pretrained general features.

\section{Material and Methods}

\begin{figure}[]
    \centering
    \includegraphics[width=0.98\linewidth]{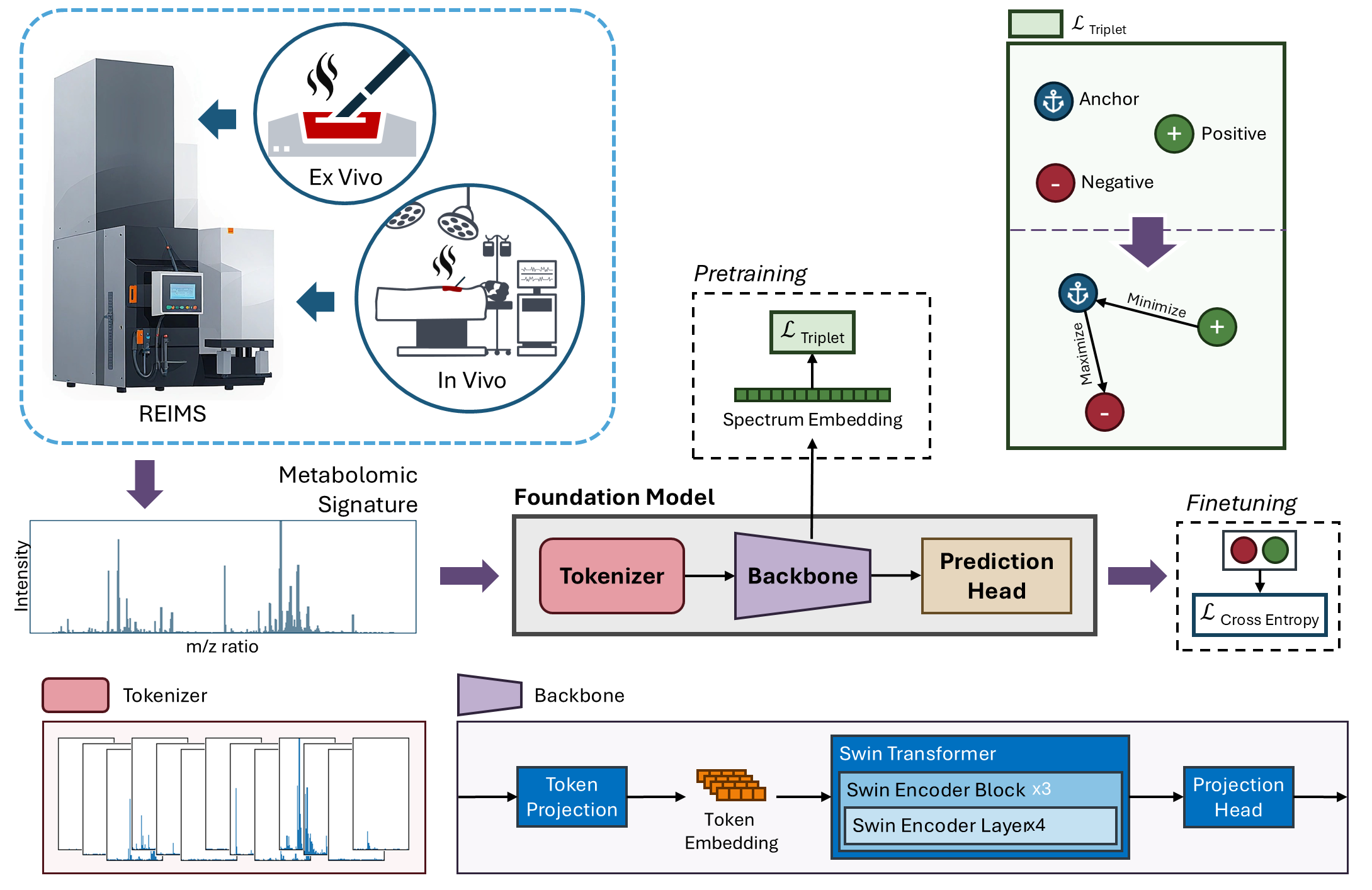}
    \caption{Overview of the FACT architecture and training process. The input spectra are tokenized, projected, and passed through a Swin Transformer encoder adapted from CLAP via transfer learning. The encoded spectrum embeddings are projected and classified using dedicated MLP heads. The model is trained in two distinct steps: pretraining using triplet loss, and finetuning using cross entropy loss.}
    \label{figure:overview}
\end{figure}

An overview of our proposed model and training strategy is illustrated in Figure~\ref{figure:overview}. REIMS analyzes the smoke generated by iKnife and outputs mass spectra at 1Hz. The spectra are then processed by our proposed model, FACT, and categorized as either cancerous or non-cancerous. FACT, is designed following the well-established architecture and widely adopted principles of transformer-based foundation models. The spectra, generated by REIMS, are broken up into \textit{tokens}, and projected with an encoder onto an embedding space, whereby they can be classified with a Multi-Layer Perceptron (MLP) head. The model is trained in two stages; first it is  \textit{pretrained} to restructure its embedding space using our proposed triplet loss method, and next it is \textit{finetuned} to accurately classify input spectra.

\subsection{Data}

The data used in this study is a proprietary collection gathered across 13 surgical clinics at the Kingston Health Sciences Center in Kingston, ON, Canada. Patients undergoing BCC resection were recruited following protocols that were approved by the institutional review board. The dataset includes both \emph{in vivo} and \emph{ex vivo} data.

\textbf{In vivo data:} During BCC resection, which takes 4–5 minutes on average, the iKnife continuously recorded mass spectra from the surgical plume at a frequency of 1 Hz, capturing real-time data throughout the procedure. In total, 10,407 spectra have been collected intraoperatively from 43 resections.

\textbf{Ex vivo data:} Following BCC resections, a dermatopathologist applied point burns on the excised tissue with a fine-tip cautery connected to the iKnife. Spectra corresponding to the burns were then labeled based on any visible pathology near that location. In total, the dataset comprises 693 samples, with 252 BCC, and 441 benign skin tissues, drawn from dermis, epidermis, or adipose of 91 patients.

\textbf{Preprocessing:} All spectra were preprocessed using native iKnife software for normalization (total ion current), calibration (lock-mass), peak binning (1 m/z intervals), and range selection focusing on the 100–1000 m/z range, as explained in~\cite{jamzad2020improved}. We employ the proposed augmentation method, termed intensity-aware augmentation, simulating calibration errors and adding distinct noise to dominant and lower-intensity peaks to expand the training data while preserving tissue-specific molecular patterns.

\subsection{Model}

We refer to the transformer encoder, along with the preceding token projection layer and the embedding projection head that follows it, as the \textit{backbone} of the model. FACT borrows the transformer encoder component found in CLAP along with its corresponding projection layers. \replaced{The encoder block is a Swin encoder \cite{liu2021swin}, a unique transformer model that employs a specialized attention mechanism called Shifted Window Attention (SWA). SWA integrates a hierarchical inductive bias, enabling the model to effectively capture both local and global dependencies while distinguishing between them—an essential capability for processing REIMS data and a crucial step in our pipeline.}{The encoder block is, in fact, a Swin encoder \cite{liu2021swin}, a unique transformer model with a special attention mechanism known as Shifted Window Attention (SWA). SWA introduces a hierarchical inductive bias to the model that can effectively capture local and global dependencies and contrast them with one another, a crucial feature for processing REIMS data.} Transfer learning from CLAP ensures that these components are well-suited for processing mass spectra, leveraging their ability to handle spectral patterns similar to audio signals. 

In order for SWA to granularly process and analyze the spectra, the spectra have to be \textit{tokenized}. We do this by partitioning the spectra into non-overlapping windows of 64-bins. This is similar to how Vision Transformers \cite{dosovitskiy2020image} process images in non-overlapping patches. The split segments corresponding to a single spectrum are then passed to the transformer as a single sequence. Individual tokens are projected onto a 96-D embedding space before being passed to the encoder as a sequence. Due to the nature of our data, we do not employ a masked attention mechanism to enforce a temporal structure to the model. Rather, the tokens comprising a single spectrum are processed as one and aggregated into a 512-D embedding vector. This embedding vector is finally classified by a simple 2-layer MLP head to predict the class of the input spectrum. This architecture allows for efficient feature extraction and classification, enabling real-time performance during deployment in surgery.

\subsection{Training}


\textbf{Pretraining:} We propose to use \textbf{triplet loss}~\cite{chechik2010large} for pretraining, a contrastive loss function for metric learning, popularized by works like FaceNet~\cite{schroff2015facenet}. The use of this loss function frames the task as a metric learning problem, where the goal is to learn an embedding space in which semantically similar samples are positioned close together, while dissimilar samples are far apart. This means that, given two spectra, the model can effectively determine their similarity by measuring the distance between their corresponding embeddings. The triplet loss function is defined as follows:

\begin{equation}
    \mathcal{L}_{\text{Triplet}} \left(a,p,n\right) = \max\{\mathcal{D}(a,p) - \mathcal{D}(a,n) + \mathrm{margin},0\}
\end{equation}

where $a$, $p$, and $n$ denote the embeddings corresponding to the \emph{anchor}, the \emph{positive} sample, and the \emph{negative} sample, respectively. The positive sample is a drawn from the same class as the anchor, while the negative sample is drawn from a different class. The loss function encourages the model to minimize the distance between the anchor and the positive sample, \(\mathcal{D}(a,p)\), while maximizing the distance between the anchor and the negative sample, \(\mathcal{D}(a,n)\). The \(\mathrm{margin}\) is a hyperparameter that enforces a minimum distance between embeddings of different classes, helping the model learn more discriminative features.

During pretraining, FACT forms a triplet with an anchor by selecting a positive spectrum from the same class and a negative spectrum from a different class. Through a technique known as online hard negative mining, the model dynamically identifies the most challenging negative samples during training—those that are closer to the anchor in the embedding space than the positive sample. This approach encourages the model to learn robust embeddings that can effectively differentiate between benign and cancerous samples, even when their spectral patterns are similar. On the other hand, triplet loss is notorious for its sensitivity to noisy labels. Yet, the well-structured embedding space created by triplet loss can be explored using conventional dimension reduction methods, to investigate and identify such problematic samples.

\textbf{Finetuning:} After pretraining, the model undergoes a finetuning phase where the learned embeddings are further refined for the specific classification task. In this stage, the embeddings are passed through the MLP prediction head, mapping the embeddings to a 2D vector, indicating the class probabilities, and the model is trained using a cross-entropy loss. This standard classification loss ensures that the model learns to accurately predict the class labels (benign or BCC) based on the embeddings generated from the pretraining stage.

\subsection{Baselines}
\label{subsection:baselines}

\textbf{Backbones:} We examine Contrastive Language-Image Pre-Training (CLIP) \cite{radford2021learning} (\(\approx~87M\) parameters) as an alternative to CLAP as the backbone. CLIP, like CLAP, is designed to associate inputs from different domains—in its case, images and text. An earlier study on mass spectrometry demonstrates that REIMS data can be effectively converted into images and classified via conventional image classifiers \cite{connolly2024imspect}. As such, we can adapt components from CLIP in the same fashion as CLAP and then pretrain and finetune the model using the same strategy.

We also evaluate Deep Representations Empowering the Annotation of Mass Spectra (DreaMS) \cite{bushuiev2024emergence} (\(\approx 118M\) parameters). DreaMS is a foundation model tailored for tandem mass spectrometry, a mass analysis technique popular in chemistry. Tandem mass spectrometry is not dissimilar to REIMS, but the process involves a \emph{fragmentation} step that simplifies distinguishing ions that have very similar m/z ratios. DreaMS's tokenizer is specifically designed to exploit fragmentation patterns within data, which are simply not present in REIMS data. Yet, as one of very few foundation models tailored to a modality similar to REIMS, it is worth examining.

\textbf{Pretraining methods:} In addition to triplet loss, we explore two alternative pretraining methods: SimCLR~\cite{chen2020simple} and FixMatch~\cite{sohn2020fixmatch}. SimCLR is a self-supervised contrastive method that brings "similar" pairs (two random augmentations of the same sample) closer in the embedding space, while pushing "dissimilar" pairs (augmentations of different samples) apart. This approach enables the use of unlabeled data, which is often more abundant. FixMatch, a semi-supervised method, combines labeled and unlabeled data by assigning high-confidence \textit{pseudo-labels} to weakly augmented samples and enforcing consistency on strongly augmented versions. These methods are particularly valuable in our study, as they allow us to leverage both the labeled ex vivo data and the significantly larger unlabeled in vivo data. By integrating the two datasets, SimCLR and FixMatch maximize the use of available data, reducing the reliance on extensive manual annotations while improving feature representation. This makes them well-suited for scenarios like ours, where labeled data is limited and unlabeled data is more abundant.

\textbf{Baselines in the Literature.} In literature, a combination of Principal Component Analysis (PCA) and Linear Discriminant Analysis (LDA) is often used to establish a linear baseline for REIMS data. In this approach, the preprocessed spectra—900-dimensional vectors following our pipeline—are first projected onto a lower-dimensional space sufficient to explain 99\% of the variance in the data. An LDA model is then fit to identify a linear decision boundary between classes. While this relatively simple method performs exceptionally well with REIMS data in certain cases, such as breast cancer detection \cite{santilli2020perioperative}, it struggles when applied to BCC data. Additionally, a 3-layer MLP with non-linear activations is commonly employed as a baseline for deep learning models. Similar to the PCA-LDA pipeline, the MLP is fed preprocessed spectra as 900-dimensional inputs and maps them to class scores. In addition to reproducing these models, we also compare our model against the results reported in \cite{fooladgar2022uncertainty,connolly2024imspect}, which explore the use of a Bayesian Neural Network (BNN) and ImSpect---an adapted image classifier--- respectively. Notably, the latter also examines SimCLR for pretraining the model.

\section{Experiments}

We evaluate the performance of FACT against the baselines discussed in \ref{subsection:baselines}. In addition to benchmarking against these methods, we visualize the embedding space of our model using Uniform Manifold Approximation (UMAP) \cite{leland2018uniform} to examine the samples that the model frequently misclassifies, and identify likely underlying issues behind these cases. We perform an extensive ablation study where we investigate the potential of several foundation models in conjunction with different pretraining methods.

\textbf{Model evaluation:} We stratify our ex vivo data following the same strategy as \cite{fooladgar2022uncertainty,connolly2024imspect}. The spectra are split into three subsets for training, validation, and testing, without patient overlap. The validation set is used for selecting the best-performing model after pretraining or finetuning, while the test set is held out for performance evaluation. For all two-stage training scenarios (pretraining followed by finetuning), the model is pretrained 30 times, and the best checkpoint is selected based on validation loss, except for triplet-loss pretraining, where silhouette score is used. This selected checkpoint is then finetuned 30 times, and performance metrics are averaged to account for variability. We calculate the balanced accuracy, sensitivity, and specificity of the model using a 0.5 decision threshold. Additionally, we report the area under the receiver operating characteristic curve (AUROC) to provide a comprehensive assessment of the model's classification accuracy across different thresholds. Performance metrics are averaged over 30 experiments for baseline comparison and 15 experiments for ablation studies, each with different random seeds, to account for variability. To assess statistical significance, metrics are compared using a one-tail Wilcoxon signed-rank test in the ablation study, as it is well-suited for paired comparisons. For benchmarking, where individual performance values are unavailable for BNN and ImSpect, we use the one-sample z-test to compare our averaged performance against reported baselines.

\textbf{Implementation details:} All experiments are conducted using Python 3.11 and PyTorch 2.1. Each experiment ran for up to 8 hours on an NVIDIA Quadro RTX 6000 (24GB VRAM). The training set was expanded using the intensity-aware augmentation method (436 → 1744) following the strategy established in \cite{fooladgar2022uncertainty} unless in cases where it adversely affected the performance. Furthermore, under CLAP and DreaMS, the same method was used for augmentations in SimCLR and FixMatch, but for CLIP, we used the pipeline presented in \cite{connolly2024imspect}. Hyperparameters were tuned via manual search based on performance on the validation set. We examined 0.25, 0.5, 1, 1.5, and 2 for \textit{margin} and found a value of 1 to perform best with triplet loss. All models were trained using AdamW \cite{loshchilov2017decoupled} with a learning rate of $10^{-4}$, $\beta_1=0.9$, $\beta_2=0.999$, a weight decay coefficient of 0.01, and a batch size of 128, except for experiments with triplet loss, which performed better with stochastic gradient descent (without momentum) with a learning rate of $10^{-1}$. Pretraining and finetuning were performed for up to 300 epochs and 25 epochs, respectively, with early stopping applied if validation loss did not improve. Full code and configurations is available on \href{https://github.com/med-i-lab/FACT/}{GitHib}.

\section{Results and Discussion}

Table \ref{table:baseline-comparison} presents the performance of our proposed approach, FACT, in comparison with the baselines. FACT achieves a balanced accuracy of \(77.5\% \pm 1.9\) and an AUROC of \(82.4\% \pm 0.8\), with the former being significantly higher than all other models (based on one-sample z-test, \(p\text{-value} < 0.001\). In comparison to prior studies, we observe that BNN outperforms FACT in terms of sensitivity, and ImSpect exceeds it in terms of specificity. However, neither model manages to surpass its balanced accuracy, indicating that FACT performs well across both classes, with less bias towards either class. On the other hand, in comparison to foundation models---without pretraining---FACT remains significantly superior, as these models do not manage to even exceed prior studies, though, the results hint at their potential. Among the three choices, CLAP performs marginally better, while CLIP is closely behind in terms of both balanced accuracy and AUROC. CLAP's better performance reinforces our hypothesis that audio Mel-spectrograms are a sufficiently close modality to REIMS data. Lastly, DreaMS falls rather short on all fronts, barely matching the linear PCA/LDA baseline. This is a strong indicator that DreaMS is not the right choice for REIMS data. Finally, FACT achieves an inference time of 72ms, which can match REIMS's maximum output rate of 1 Hz for real-time application in intraoperative settings. This demonstrates the practical viability of the model, alongside its superior classification accuracy.

\begin{table}[h!]
    \caption{Performance of our proposed approach in comparison to baselines.\\Values show the mean and the standard deviation of each metric under 30 runs.}
    \label{table:baseline-comparison}
    \vspace{-10pt}
    \begin{tabular}{@{}llllll@{}}
        \toprule
        Model & Pretraining & \begin{tabular}[c]{@{}l@{}}Balanced\\Accuracy\end{tabular} & Sensitivity & Specificity & AUROC \\
        \midrule
        PCA/LDA & - & 69.7\% ± 2.3 & 66.0\% ± 3.3 & 73.5\% ± 1.7 & 73.9\% ± 1.3 \\
        MLP & - & 73.9\% ± 2.7 & 67.8\% ± 6.1 & 79.9\% ± 5.3 & 79.9\% ± 1.9 \\
        BNN \cite{fooladgar2022uncertainty} & - & 75.2\% ± 2.9 & 74.1\% ± 3.8 & 77.3\%   ± 4.8 & 82.1\% ± 2.6 \\
        ImSpect \cite{connolly2024imspect} & \checkmark & 73.5\% ± 1.4 & 63.1\% ± 2.6 & \textbf{83.9\% ± 1.8} & 81.6\% ± 1.4 \\
        \midrule
        CLAP \cite{wu2023large} & - & 71.9\% ± 1.6 & 65.8\% ± 5.0 & 78.0\% ± 5.3 & 78.1\% ± 2.3 \\
        CLIP \cite{radford2021learning} & - & 69.8\% ± 2.9 & 59.8\% ± 10.6 & 81.1\% ± 9.9 & 75.8\% ± 2.9 \\
        DreaMS \cite{bushuiev2024emergence} & - & 68.2\% ± 2.7 & 61.8\% ± 6.2 & 74.6\% ± 5.8 & 73.7\% ± 2.2 \\
        \midrule
        FACT \textit{(ours)} & \checkmark & \textbf{77.5\% ± 1.9} & 72.2\% ± 4.3 & 82.8\% ± 1.6 & \textbf{82.4\% ± 0.8} \\
        \botrule
    \end{tabular}
    \vspace{-10pt}
\end{table}

\begin{figure}[t]
    \centering
    \includegraphics[width=1\linewidth]{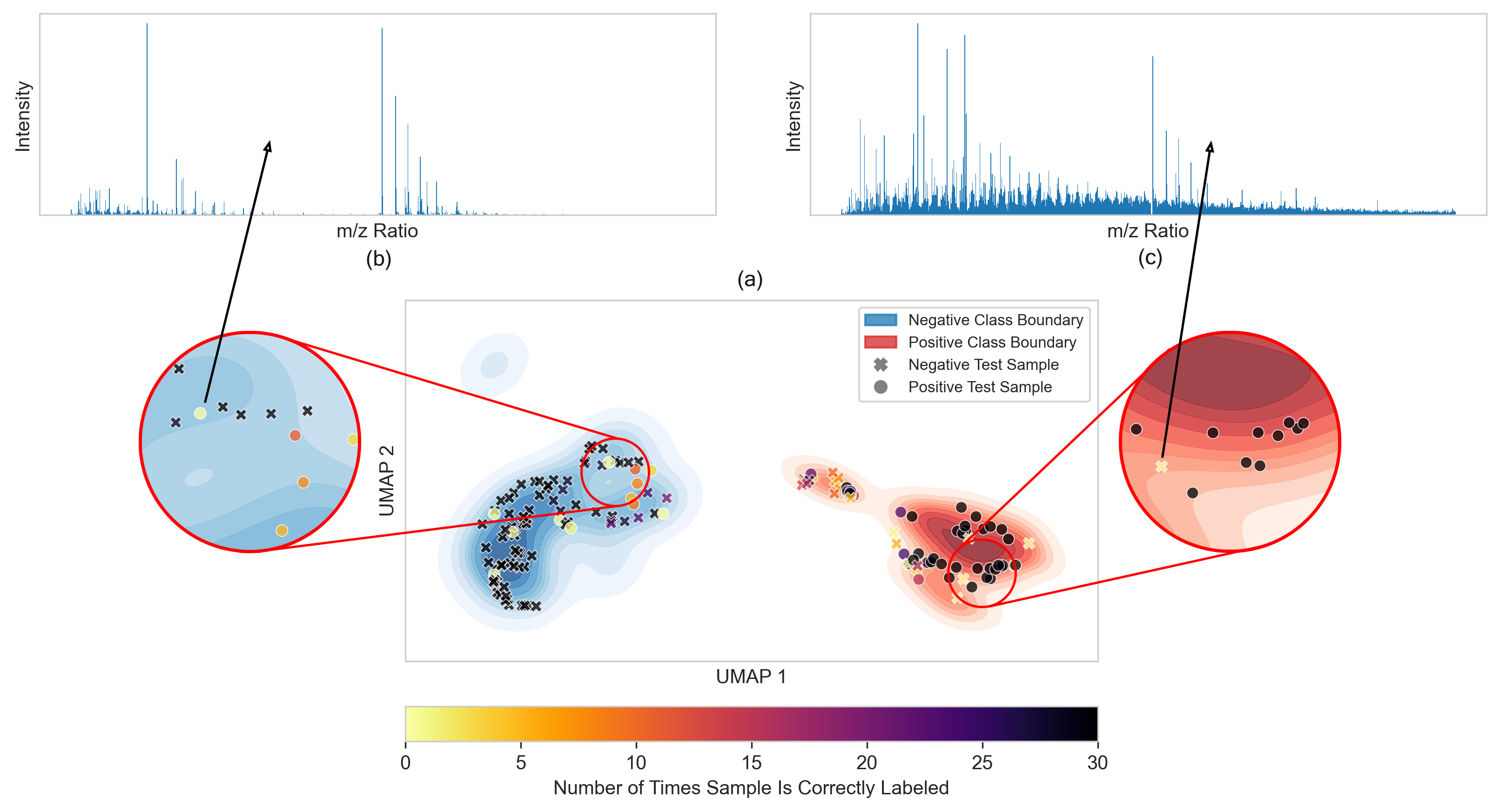}
    \vspace{-10pt}
    \caption{(a) The embedding space of FACT reduced to 2 dimensions using UMAP. Decision boundaries of the classes are visualized as contours using the training samples, while test samples are displayed as points, with crosses depicting negative samples and circles designating positive samples. Samples are colored based on how frequently they are correctly classified. Most samples align with the decision boundaries, and are thus, shown in a darker shade, but a few samples, often near the edges, are misclassified. (b) and (c) are sample spectra with high failure rate drawn from the test set, which are labeled positive and negative, respectively.}
    \label{figure:failure-modes}
\end{figure}

\textbf{Failure modes:} Figure \ref{figure:failure-modes}a visualizes the embedding space of our model after pretraining.  As illustrated, the majority of samples are correctly positioned within the expected class boundaries, forming two distinct clusters predominantly comprising positive and negative training samples. Misclassified test samples tend to appear near the periphery of the incorrect bounds. Notably, the likelihood of misclassification increases for samples situated deeper within the boundaries of the opposing class. This observation suggests that the model struggles to classify such instances, indicating that these cases are more ambiguous or present characteristics that align more closely with the incorrect class. A closer examination of such misclassified samples reveals a correlation with sample noise. Figure \ref{figure:failure-modes}b-c exemplifies this issue, displaying two spectra from the test set with positive and negative labels, respectively. The spectrum in Figure~\ref{figure:failure-modes}b lacks any discernible tissue-related signature, as indicated by the distribution of ions in fatty acids ($\sim$ 150-400 m/z) and glycerophospholipids ($\sim$ 600-900 m/z) ranges. This is most likely due to suboptimal data acquisition settings. In contrast, the failure of the sample in Figure~\ref{figure:failure-modes}c is likely attributed to a low signal-to-noise ratio, resulting from high background noise. In addition to technical factors, it is important to acknowledge the potential for biological uncertainty due to the diffuse nature of cancerous cells. While the ex vivo data collection minimized this uncertainty by sampling from homogeneous regions under the guidance of a histopathologist, some degree of label noise remains unavoidable.

\begin{figure}[h!]
    \centering
    \includegraphics[width=1\linewidth]{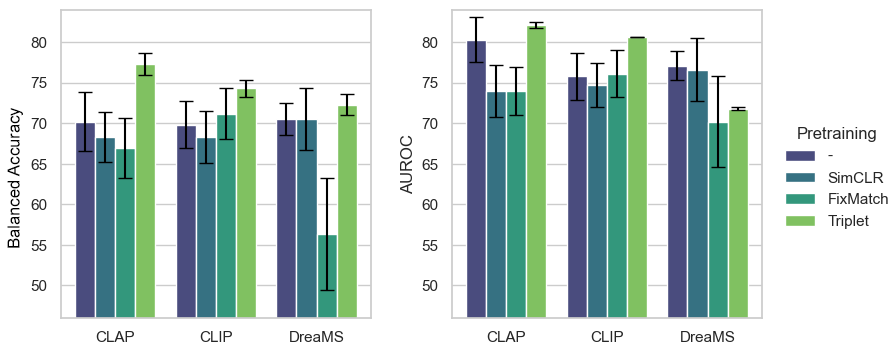}
    \vspace{-5pt}
    \caption{Performance of foundation models under different pretraining methods.\\The figure illustrates the mean and the standard deviation of each metric under 15 runs.}
    \label{figure:ablation-study}
\end{figure}

\textbf{Ablation study:} Last but not least, the results of our ablation studies regarding the candidate backbones and pretraining methods are presented in Figure \ref{figure:ablation-study}. CLAP and Triplet, the building blocks of FACT, show the best overall performance with a balanced accuracy of 77.9\% ± 1.2 and an AUROC of 84.8\% ± 0.8, both of which are significantly better than other results (based on one-tail Wilcoxon test, \(p\text{-value} < 0.05\)). The pairing also achieves a high specificity score of 80.7\% ± 5.1 without sacrificing sensitivity, like CLIP. On the other hand, CLIP, which achieves its best results without pretraining, also shows the best performance under self-supervised and semi-supervised pretraining methods. CLIP's advantage in this regard can be ascribed to the augmentation methods used in SimCLR and FixMatch. Extensive literature on weakly supervised learning in vision has established a versatile augmentation pipeline applicable to many problems. But the work on mass spectrometry, and specially REIMS, is in early stages, and alternatives to intensity-aware augmentation can lead to significant improvements with SimCLR and FixMatch. Finally, DreaMS degrades in performance under self-supervised and semi-supervised methods and its overall behavior is highly variable and inconsistent, as evident by the high standard deviations. The fault here likely lies with DreaMS tokenization approach, which is specifically designed for tandem mass spectrometry and to exploit fragmentation patterns, which are simply not present in REIMS. Overall, these results reinforce the effectiveness of our triplet pretraining, particularly when paired with the CLAP backbone, in achieving a balanced and accurate classification.

\section{Conclusion}

In this study, we presented FACT, a novel foundation model for the classification of cancer tissue margins using REIMS and iKnife. By adapting the architecture of CLAP and implementing a supervised pretraining method based on triplet loss, we demonstrated significant improvements over existing approaches, achieving balanced accuracy and AUROC scores that surpassed previous baselines. Our results emphasize the effectiveness of foundation models, particularly when tailored for specific modalities, in enhancing the classification of complex mass spectrometry data, even with limited labeled examples. Our analysis also highlighted certain failure modes, notably the model's difficulty in correctly classifying noisy samples, i.e. spectra lacking tissue-related signatures or those with low signal-to-noise ratios. These findings suggest that suboptimal data acquisition and high background noise contribute to model errors, underscoring the need for more robust preprocessing and data quality control.

\textbf{Future Work:} Given the multi-modal nature of CLAP (and CLIP) there is potential to extend this approach by integrating text prompts with spectra data to create a more versatile and robust model. By leveraging textual descriptions of samples or clinical contexts, future iterations of FACT could provide even more accurate and context-aware classification, paving the way for a deeper understanding of tissue characterization during surgery. Exploring alternative tokenization approaches, which would also involve more sophisticated sequential modeling approaches, could further enhance the robustness and performance of the model. Additionally, \added{with state-space models (SSMs) emerging as a new wave in foundation models for sequential data, investigating their applicability to REIMS data could provide new insights into spectral representation. Finally,} as a foundation model, FACT could be explored on other downstream tasks beyond classification. Tasks such as anomaly detection, regression for quantitative tissue properties, or segmentation of spectral regions could further demonstrate the versatility and generalizability of the proposed approach.

\section*{Declarations}

\textbf{Compliance with Ethical Standards:} This study was approved by Queen’s University Health Sciences Research Ethics Board.

\noindent
\textbf{Funding:} This work is supported by the Natural Sciences and Engineering Research Council, the Canadian Institutes of Health Research, Canada Research Chair (Gabor Fichtinger, Parvin Mousavi), Surgical Innovation Chair (John F. Rudan), and Canada CIFAR AI chair and the Vector Institute (Parvin Mousavi).

\noindent
\textbf{Conflict of Interest:} We have no conflicts of interest to declare.

\noindent
\textbf{Informed Consent:} All patients provided informed verbal and written consent.

\noindent
\textbf{Data \& Code Availability:} The data from this study is not available. The code is available on \href{https://github.com/med-i-lab/FACT/}{GitHib}.

\bibliography{bibliography}

\end{document}